\documentclass[fleqn,10pt]{wlscirep}
\usepackage{graphicx}
\usepackage{amsmath}
\usepackage{amssymb}
\usepackage{bm}
\usepackage{epstopdf}
\usepackage{epsfig}
\usepackage{color}
\usepackage{caption}

\def\A{\uparrow}
\def\V{\downarrow}

\title{Triplet $p$-wave pairing correlation in low-doped zigzag graphene nanoribbons}
\author[1,2,*]{Tianxing Ma}
\author[3,\dag]{Fan Yang}
\author[4,2,\ddag]{Zhongbing Huang}
\author[2]{Hai-Qing Lin}
\affil[1]{Department of Physics, Beijing Normal University,
Beijing 100875, China}
\affil[2]{Beijing Computational Science Research Center,
Beijing 100193, China}
\affil[3]{School of Physics, Beijing Institute of Technology,
Beijing 100081, China}
\affil[4]{Faculty of Physics and Electronic Technology, Hubei University,
Wuhan 430062, China}

\affil[*,\dag,\ddag]{Correspondence and requests for materials should be addressed to T. Ma* (txma@bnu.edu.cn), F. Yang\dag (yangfan\_blg@bit.edu.cn) or Z. Huang\ddag (huangzb@hubu.edu.cn).}
 
\begin{abstract}
We reveal an edge spin triplet $p-$wave superconducting pairing correlation in slightly doped zigzag graphene nanoribbons. By employing a method that combines random-phase approximation, the finite-temperature determinant quantum Monte Carlo approach, and the ground-state constrained-path quantum Monte Carlo method, it is shown that such a spin-triplet pairing is mediated by the ferromagnetic fluctuations caused by the flat band at the edge.
The spin susceptibility and effective pairing interactions at the edge strongly increase as the on-site Coulomb
interaction increases, indicating the importance of electron-electron correlations.
It is also found that the doping-dependent ground-state $p$-wave pairing correlation bears some similarity to the famous
superconducting dome in the phase diagram of a high-temperature superconductor, while the spin correlation at the edge
is weakened as the system is doped away from half filling.
\end{abstract}
\begin{document}

\flushbottom
\maketitle
%
%
\thispagestyle{empty}


\section*{Introduction}

Triplet superconductivity (SC) has been a focus of modern condensed matter physics because of its possible connection to topological quantum information
and computation\cite{Kitaev2001,Alicea2012,PhysRevLett.107.217001,Mourik1003,Deng2012,Rokhinson2012,PhysRevLett.109.056803,Anindya2012,PhysRevB.87.241401,PhysRevA.82.053611}. It has been proposed that a gapless Majorana bound state would localize at the end of the one-dimensional spinless $p-$wave superconductor\cite{Kitaev2001}, which could be used to practically realize topological quantum computation\cite{Kitaev2003,RevModPhys.80.1083}. To realize such a Majorana bound state in real material, the superconducting proximity effect was proposed \cite{PhysRevLett.105.077001,PhysRevLett.106.127001,PhysRevLett.105.227003}, and experimental evidence of its existence was recently reported\cite{Perge602}. Here, we explore the possibility of intrinsic triplet SC, which is generated by an electronic correlation.

In this paper, we reveal a possible edge-spin triplet $p$-wave superconducting pairing correlation in slightly doped zigzag graphene nanoribbons with appropriate interactions.
Graphene, a single layer of carbon, has generated immense interest ever since its experimental discovery\cite{Novoselov666,RevModPhys.81.109}.
Recently, experimental advances in doping methods have made it possible to change the type of
carriers (electrons or holes)\cite{PhysRevLett.104.136803,PhysRevLett.105.256805}, opening the doors for
exotic phases, such as SC and magnetism induced by repulsive interactions. For instance, it was shown by the two-stage renormalization-group calculation that unconventional SC is induced by weak repulsive interactions in honeycomb Hubbard models that are away from half-filling\cite{PhysRevB.81.224505}, and that a topological $d+id$ SC is induced in a heavily doped system\cite{PhysRevB.75.134512,PhysRevB.78.205431,PhysRevB.81.085431,PhysRevB.86.020507,PhysRevB.84.121410,PhysRevB.85.035414,Nandkishore2012}.
At graphene edges the density of states may be peaked
due to the presence of edge-localized states close to the
Fermi level\cite{PhysRevLett.106.226401}. Especially at extended zigzag edges this
leads to a phenomenon called edge magnetism, for which various
theories \cite{PhysRevB.80.245436,PhysRevB.91.075410,LiJPCM2016} 
predict ferromagnetic (FM) intraedge and antiferromagnetic (AFM)
interedge correlations. {\color{blue} The proposed magnetism is similar to
the flat-band ferromagnetism appearing in the orbital-active optical honeycomb lattice\cite{PhysRevA.82.053618},
where the band flatness dramatically amplifies the interaction effect, driving the ferromagnetic transition even with a relatively weak repulsive
interaction}. From these discoveries, a question which naturally arises: is there is triplet SC mediated by the FM spin correlations on each edge in the doped zigzag graphene nanoribbons?

\begin{figure}[tbp]
\begin{center}
\includegraphics[scale=0.45]{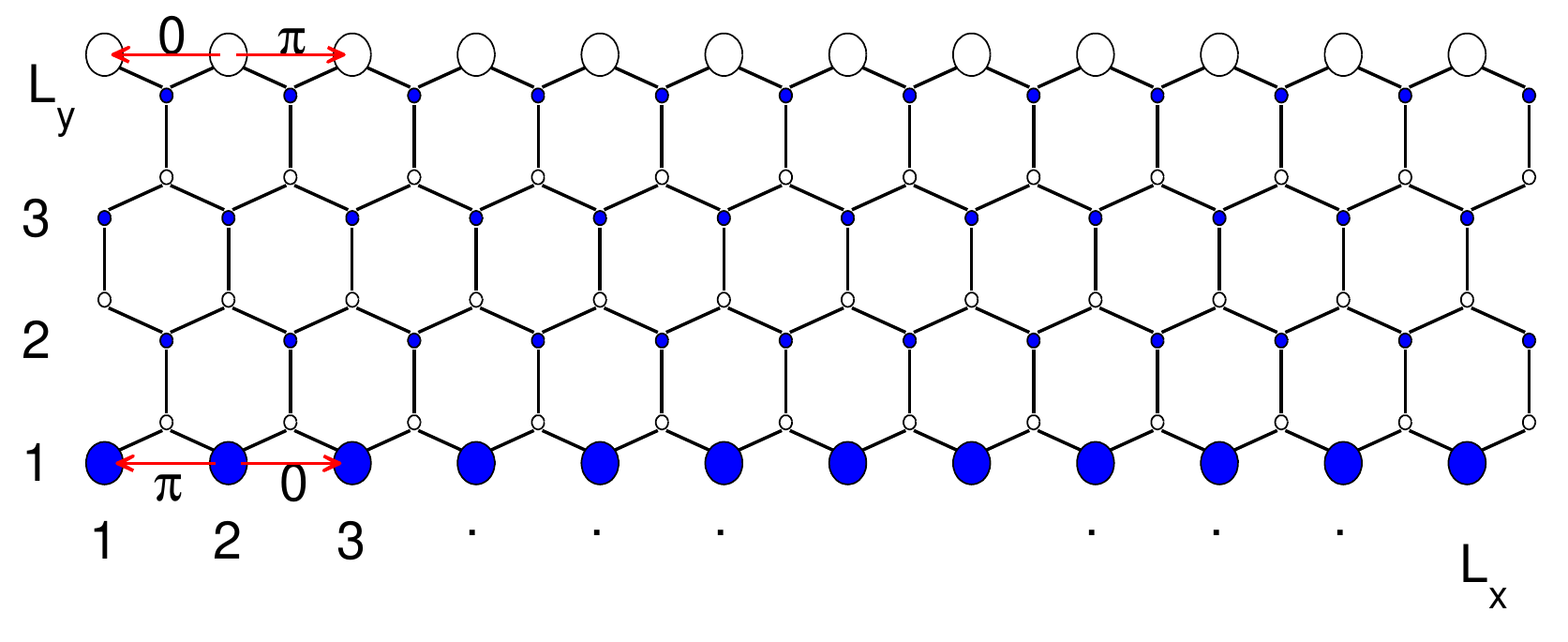}
\caption{(Color online) A piece of a honeycomb lattice displaying zigzag edges with $L_y=4$ which defines the width
of the ribbon in the transverse direction and $L_x$=12, which defines the length in the longitudinal direction.
The lattice sites at zigzag edge are much larger than the sites in the bulk, indicating that the charge carriers are moving along the edge.
}\label{Fig:Structure}
\end{center}
\end{figure}

\begin{figure}[tbp]
\begin{center}
\includegraphics[scale=0.4]{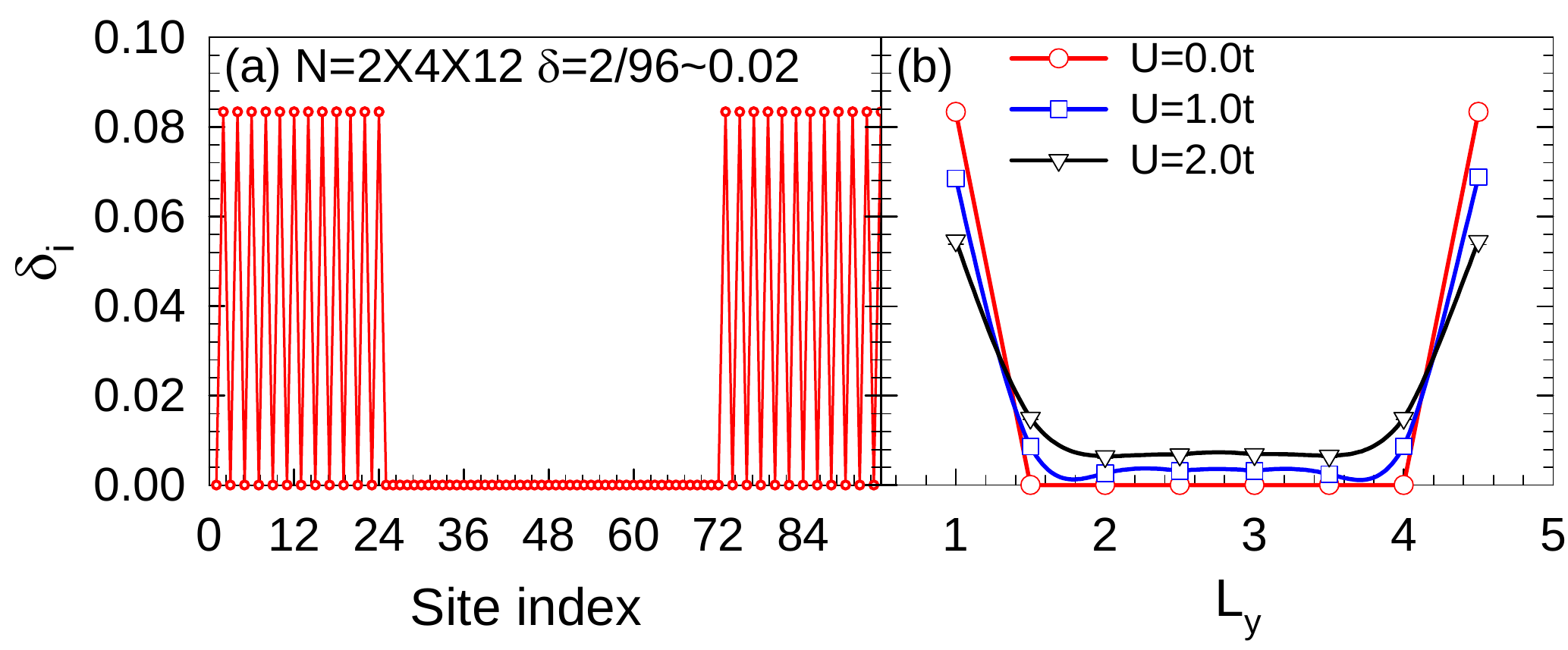}
\caption{(Color online) The carrier distribution (a) as a function of the site index at $U=2.0t$ and (b) from edge $\rightarrow$ bulk $\rightarrow$ edge with different $U$. It is clear to see that most charge carriers are distributed along the edge.
}\label{Fig:ndistrubution}
\end{center}
\end{figure}

\begin{figure}[tbp]
\begin{center}
\includegraphics[scale=0.4]{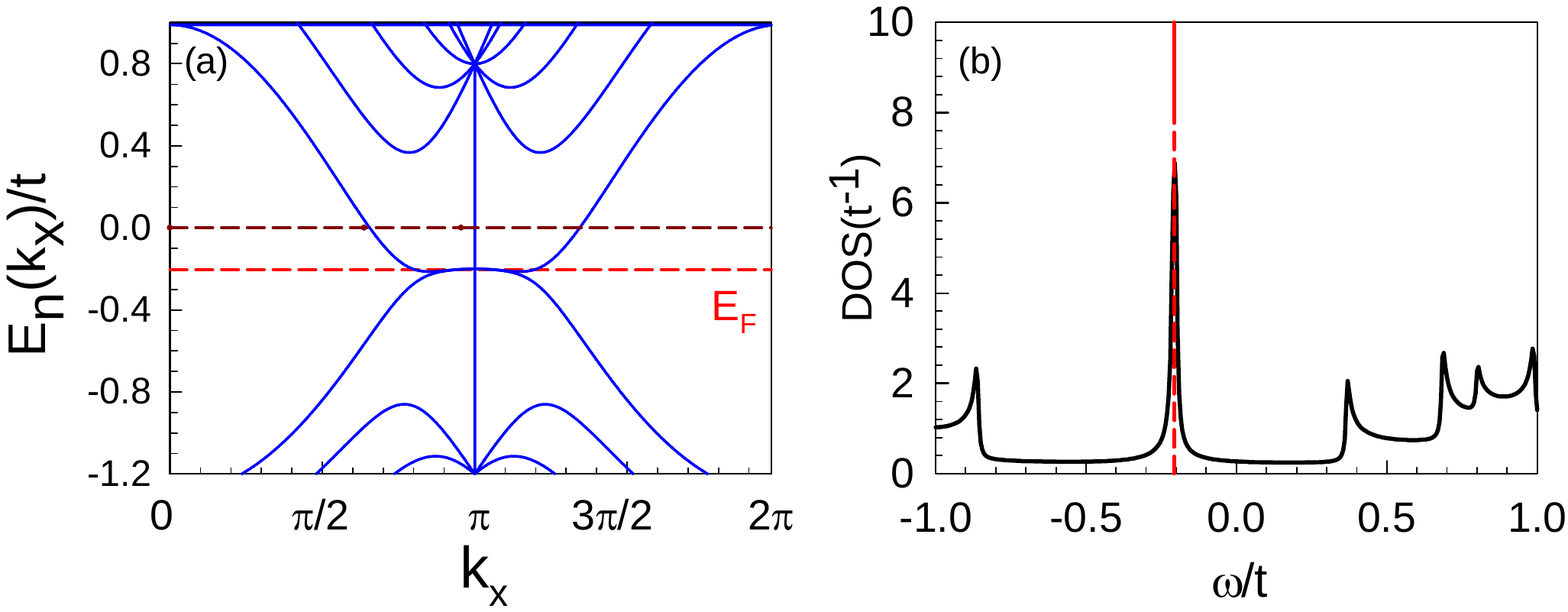}
\caption{(Color online) Band structure (a) and DOS (b) of a six-chain nanoribbon system. Note that the flat band bottom, located at approximately $-0.2t$ in (a), leads to the DOS peak in (b). The Fermi level of the half-filled system is marked by the red dashed lines in both figures. }\label{Fig:band}
\end{center}
\end{figure}
In the present work, we establish the $p$-wave superconducting pairing correlation at the edges of zigzag graphene nanoribbons by using combined random-phase approximation (RPA)\cite{RevModPhys.84.1383,PhysRevB.69.104504,JPSJ,Graser2009,PhysRevB.75.224509,PhysRevLett.101.087004,PhysRevLett.111.066804,Srep0820}, the finite-temperature determinant quantum Monte Carlo (DQMC)\cite{PhysRevD.24.2278,PhysRevB.31.4403,MaAPL2010,PhysRevLett.110.107002,PhysRevB.94.075106} and the ground-state constrained-path quantum Monte Carlo (CPQMC)\cite{PhysRevLett.74.3652,PhysRevB.55.7464,PhysRevB.84.121410,WuEPL2013,MaEPL2015} methods.
Our unbiased results show that both the ferromagnetic spin correlation and the effective $p$-wave superconducting pairing correlation are greatly enhanced as the interaction increases.

\section*{Results}

The ribbon geometry considered here is depicted in Fig. \ref{Fig:Structure}, in which the blue and white circles represent sublattices A and B, respectively, and the transverse integer index $1,2, . . . ,L_y$ defines the width of the ribbon while $1,2, . . . ,L_x$ at
the zigzag edge defines the length. 
Assuming the ribbon to be infinite in the $x$ direction but finite in the $y$ direction, we produce a graphene nanoribbon with zigzag edges.
In the following studies, the interaction $U$ is introduced through the standard Hubbard model. In Fig.\ref{Fig:ndistrubution}, the carrier distribution (a) as a function of the site index at $U=2.0t$ and (b) from edge $\rightarrow$ bulk $\rightarrow$ edge with different interactions is shown. It is clear to see that most charge carriers distribute along the edge, and the increasing interaction pushes many more charge carriers to the edges.

The band structure of a six-chain nanoribbon system is shown in Fig. \ref{Fig:band}(a).
Here, as the system is periodic in the $x$-direction, the momentum $k_x$ is a good quantum number.
From Fig. \ref{Fig:band}(a), one finds a flat band bottom with energies located near the Fermi level ($\approx-0.2t$) of the half-filled system. Physically, such a flat band bottom is caused by the edge states, which leads to the DOS peak at approximately $-0.2t$ shown in Fig. \ref{Fig:band}(b).
\begin{figure}[tbp]
\begin{center}
\includegraphics[scale=0.4]{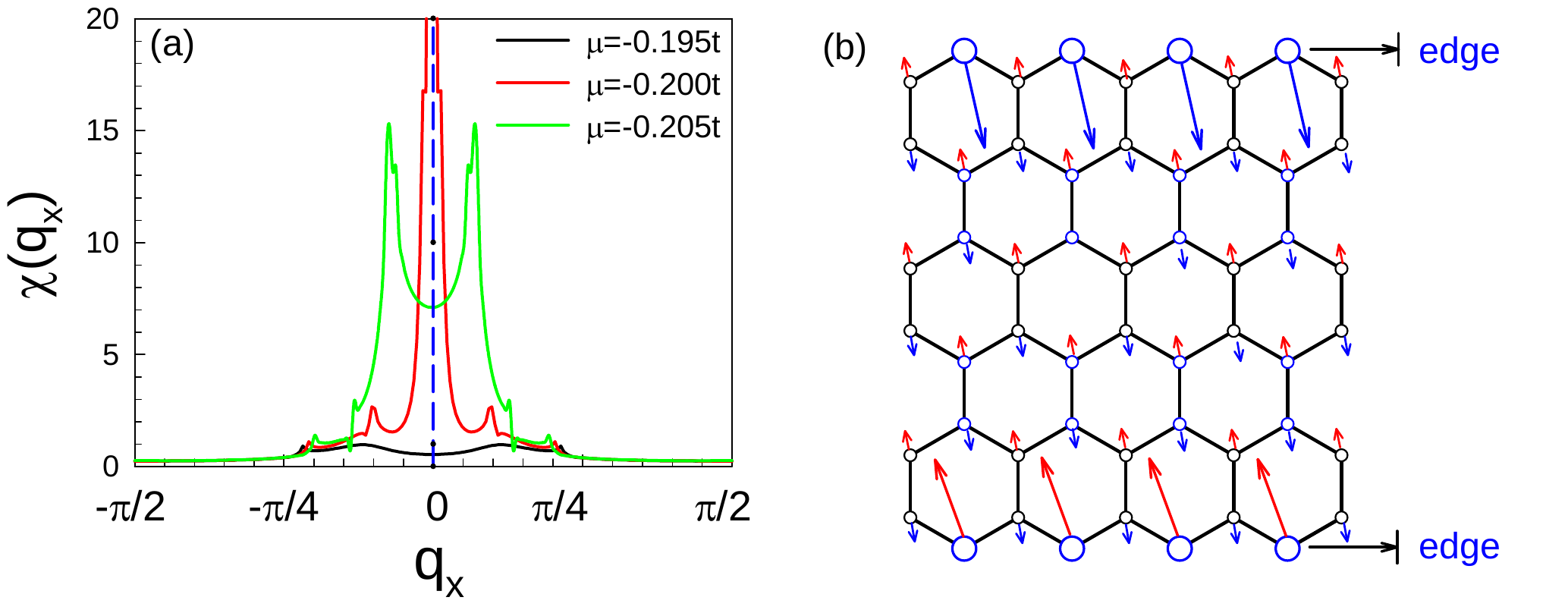}
\caption{(Color online) (a) The largest eigenvalue $\chi(q_x)$ of the susceptibility matrix $\chi^{(0)}_{l,m}\left(q_x\right)$ as a function of $q_x$  for three different dopings, i.e., $\mu=-0.195t$ ($\delta=3.6\%$), $\mu=-0.2t$ ($\delta=3.0\%$) and $\mu=-0.205t$ ($\delta=0.8\%$) for the 6-chain system near half-filling. (b) Sketch of the pattern of the dominating spin fluctuations for $\mu=-0.2t$, as determined by the eigenvector of $\chi^{(0)}_{l,m}\left(q_x=0\right)$ corresponding to its largest eigenvalue.
}\label{Fig:magnetic}
\end{center}
\end{figure}

\subsection*{RPA study} Guided by the idea that triplet SC may be mediated by the strong FM spin fluctuations in the system, we performed an RPA-based study on the
possible pairing symmetries of the system. The multi-orbital RPA approach\cite{RevModPhys.84.1383,PhysRevB.69.104504,JPSJ,Graser2009,PhysRevB.75.224509,PhysRevLett.101.087004,PhysRevLett.111.066804,Srep0820}, which is a standard and effective approach for the case of the weak coupling limit, is applied in our study for small $U$($<0.01t$). Various bare susceptibilities of this system are defined as
\begin{eqnarray}
 \chi^{(0)l_{1},l_{2}}_{l_{3},l_{4}}\left(q_x,\tau\right)\equiv
 \frac{1}{N}\sum_{k_{1},k_{2}}\left<T_{\tau}c^{\dagger}_{l_{1}}(k_{1},\tau)
 c_{l_{2}}(k_{1}+q_x,\tau)c^{+}_{l_{3}}(k_{2}+q_x,0)c_{l_{4}}(k_{2},0)\right>_0,\label{free_sus}
 \end{eqnarray}
 where $l_{i}$ $(i=1,2L_y)$ denote orbital (sublattice) indices.

The largest eigenvalue $\chi(q_x)$ of the susceptibility matrix $\chi^{(0)}_{l,m}\left(q_x\right)\equiv
\chi^{(0)l,l}_{m,m}\left(q_x,i\nu=0\right)$ as function of $q_x$ is shown in
Fig. \ref{Fig:magnetic}(a) for three different dopings near half-filling. As a result, the susceptibility for the doping $\delta=3\%$ with chemical potential  $\mu=-0.2t$ peaks at zero momentum, which suggests strong FM intra-sublattice spin fluctuations in the system. Further more, from the eigenvector of the susceptibility matrix, one can obtain the pattern of the dominating spin fluctuation in the system, which is shown in Fig. \ref{Fig:magnetic}(b). Obviously, the dominating spin fluctuation, which is mainly located on the two edges, is FM on each edge and AFM between the two edges. When $\mu$ deviates from $-0.2t$, the susceptibility peaks deviate from zero, as shown in Fig. \ref{Fig:magnetic}(a), suggesting weaker FM spin fluctuations in the system.

With weak-Hubbard $U$, the spin ($\chi^{s}$)
and charge ($\chi^{c}$) susceptibilities in the RPA level are given by
\begin{equation}
\chi^{s\left(c\right)}\left(\mathbf{q},i\nu\right)=
\left[I\mp\chi^{(0)}\left(\mathbf{q},i\nu\right)\bar U\right]^{-1}\chi^{(0)}
\left(\mathbf{q},i\nu\right),\label{RPA}
\end{equation}
where $\bar U^{\mu\nu}_{\mu'\nu'}$ ($\mu,\nu=1,\cdots, 2L_y$) is a $4L_{y}^2\times4L_{y}^2$ matrix, whose nonzero
elements are $\bar U^{\mu\mu}_{\mu\mu}=U$ ($\mu=1,\cdots,2L_y$). Clearly, the repulsive $U$ suppresses $\chi^{c}$ but enhances $\chi^{s}$. Thus, the spin fluctuations take the main role of mediating the Cooper pairing in the system. In the RPA level, via exchanging the spin fluctuations represented by the spin susceptibilities, the Cooper pairs near the FS acquire an effective interaction $V_\textrm{eff}$\cite{RevModPhys.84.1383,PhysRevLett.111.066804}, from which one solves the linearized gap equation to obtain the leading pairing symmetry and its critical temperature $T_c$.

\begin{figure}[tbp]
\begin{center}
\includegraphics[scale=0.4]{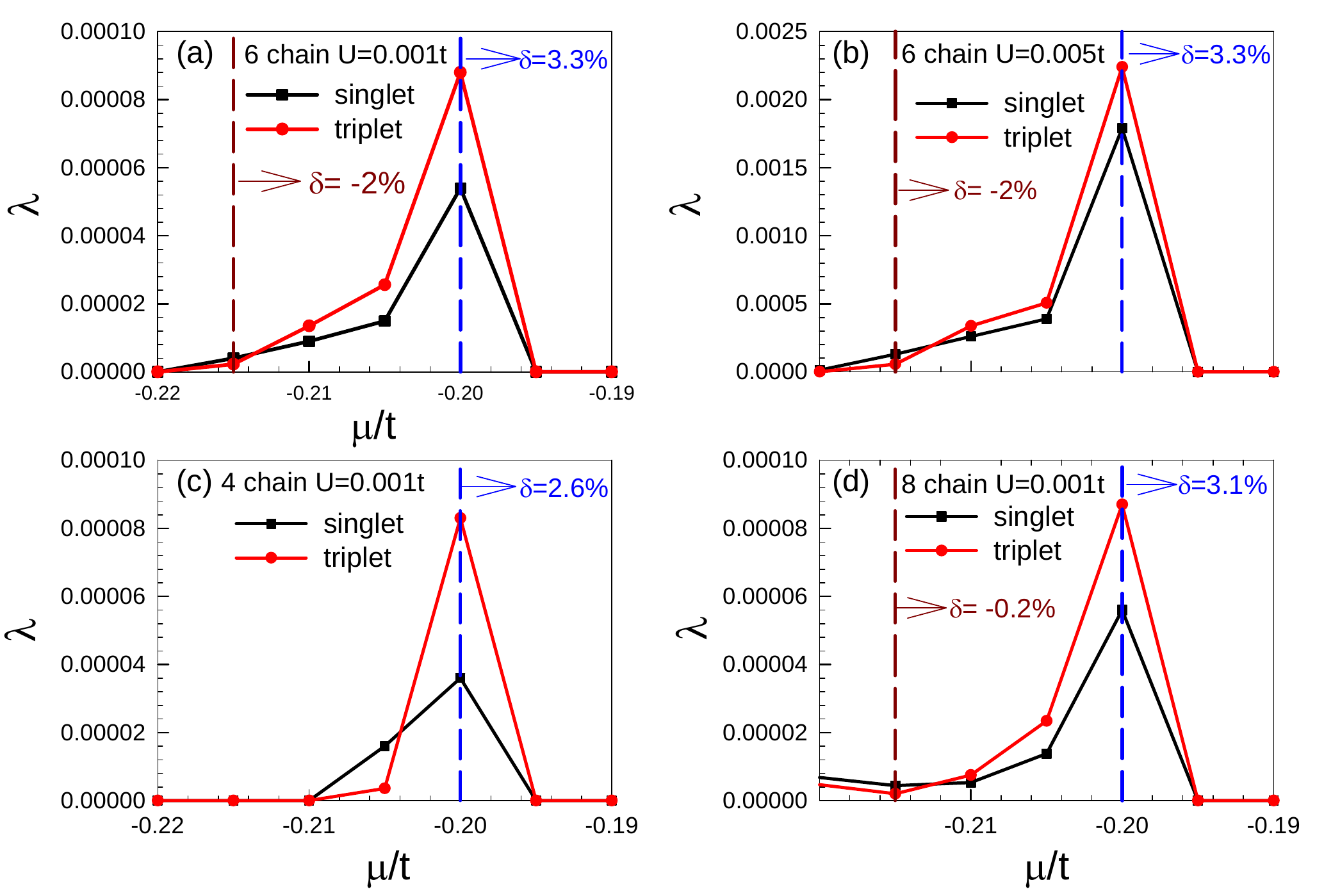}
\caption{(Color online) Doping dependence of the largest eigenvalues $\lambda$ of singlet and triplet pairings for (a)  $U=0.001t$, and (b) $U=0.005t$ for the 6-chain system, (c) $U=0.005t$ for the 4-chain system and (d) $U=0.001t$ for the 8-chain system.}\label{Fig:SC}
\end{center}
\end{figure}

Focusing on the low-doping regime in which the chemical potential $\mu$ is located within the flat band bottom, we obtained the largest pairing eigenvalues $\lambda$ for the singlet and triplet pairings as functions of $\mu$ for a 6-chain ribbon with weak interaction $U=0.001t$, as shown in Fig. \ref{Fig:SC}(a). Interestingly, while both pairings attain their largest eigenvalues at $\mu=-0.2t$ (3\% doping) due to the DOS peak there (as shown in Fig. \ref{Fig:band}(b)), the triplet pairing wins over the singlet one in the low doping regime at the flat band bottom. Physically, the triplet pairing in this regime is mediated by the FM spin fluctuations shown in Fig. \ref{Fig:magnetic}. In Fig. \ref{Fig:SC}(b), the results for $U=0.005t$ are shown. Comparing (a) and (b), it's obvious that stronger interaction leads to pairing correlations that are qualitatively the same as but quantitatively stronger than weak interaction. In Fig. \ref{Fig:SC}(c) and (d), the results for a 4-chain ribbon and 8-chain ribbon are shown with $U=0.001t$. The results for all these cases are qualitatively similar.

Note that we have chosen a very weak $U$ in our RPA calculations, since for $U>U_c\approx 0.007t$ (for $\mu=-0.2t$), the divergence of the spin-susceptibility invalidates our calculations. Physically, such a spin susceptibility divergence will not lead to a magnetically ordered state since the Mermin and Wagner's theorem prohibits a one-dimensional system from forming long-range order. Instead, short-ranged FM spin correlations here might mediate triplet superconducting pairing correlations. We leave the study of the case of $U>U_c$ to the following DQMC and CPQMC approaches, which are suitable for strong coupling problems.

\subsection*{QMC Result}
\begin{figure}[tbp]
\begin{center}
\includegraphics[scale=0.4]{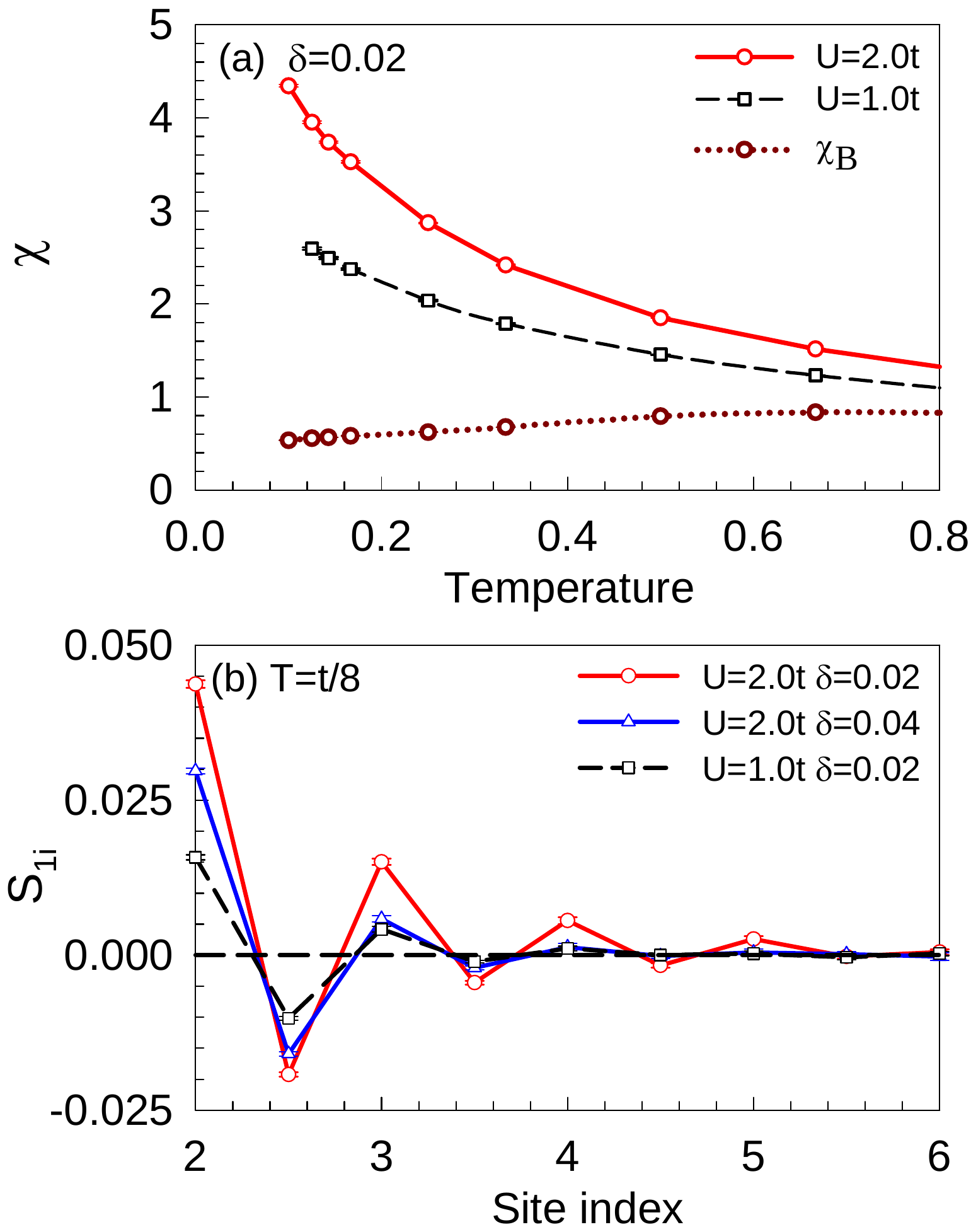}
\caption{(Color online) (a) The edge $\chi$ as a function of temperature at $\delta$=0.02 for different $U$, and the uniform $\chi_B$ for $U=2.0t$ is also present.
(b) The spin correlation as a function of the site index along the edge for $U=2.0t$ at $\delta$=0.02 and $\delta$=0.04, and $U=1.0t$ at $\delta$=0.02.
}\label{Fig:Spin}
\end{center}
\end{figure}

As FM 
fluctuations play an essential role, we first study the
magnetic correlations. In Fig. \ref{Fig:Spin} (a), the edge spin susceptibility $\chi$ is shown as a function of temperature with different $U$ at $\delta$=0.02. The edge $\chi$ is calculated by summing over the sites on the edge, such as those marked as larger circles in Fig. \ref{Fig:Structure}. It is interesting to see that $\chi$ increases as the temperature decreases, which indicates a dominant FM fluctuations on the zigzag edge. Additionally, $\chi$ increases as $U$ increases, indicating that the electronic correlation is important for the magnetic excitation in such a system. The uniform spin susceptibility $\chi_B$ for the whole system is also shown, which decreases slightly as the temperature decreases. To further reveal the FM correlation on the zigzag edge, the spin-spin correlation along the edge is shown in Fig. \ref{Fig:Spin} (b). One can see that the spin correlation $S_{1i}(i=2,3,\cdots)$ along the edge is always positive, 
suggesting FM correlation. One may also see that the spin correlation is weakened as the system is doped away from half filling, which is in agreement with the results indicated by RPA shown in Fig. \ref{Fig:magnetic}(a).

\begin{figure}[tbp]
\begin{center}
\includegraphics[scale=0.4]{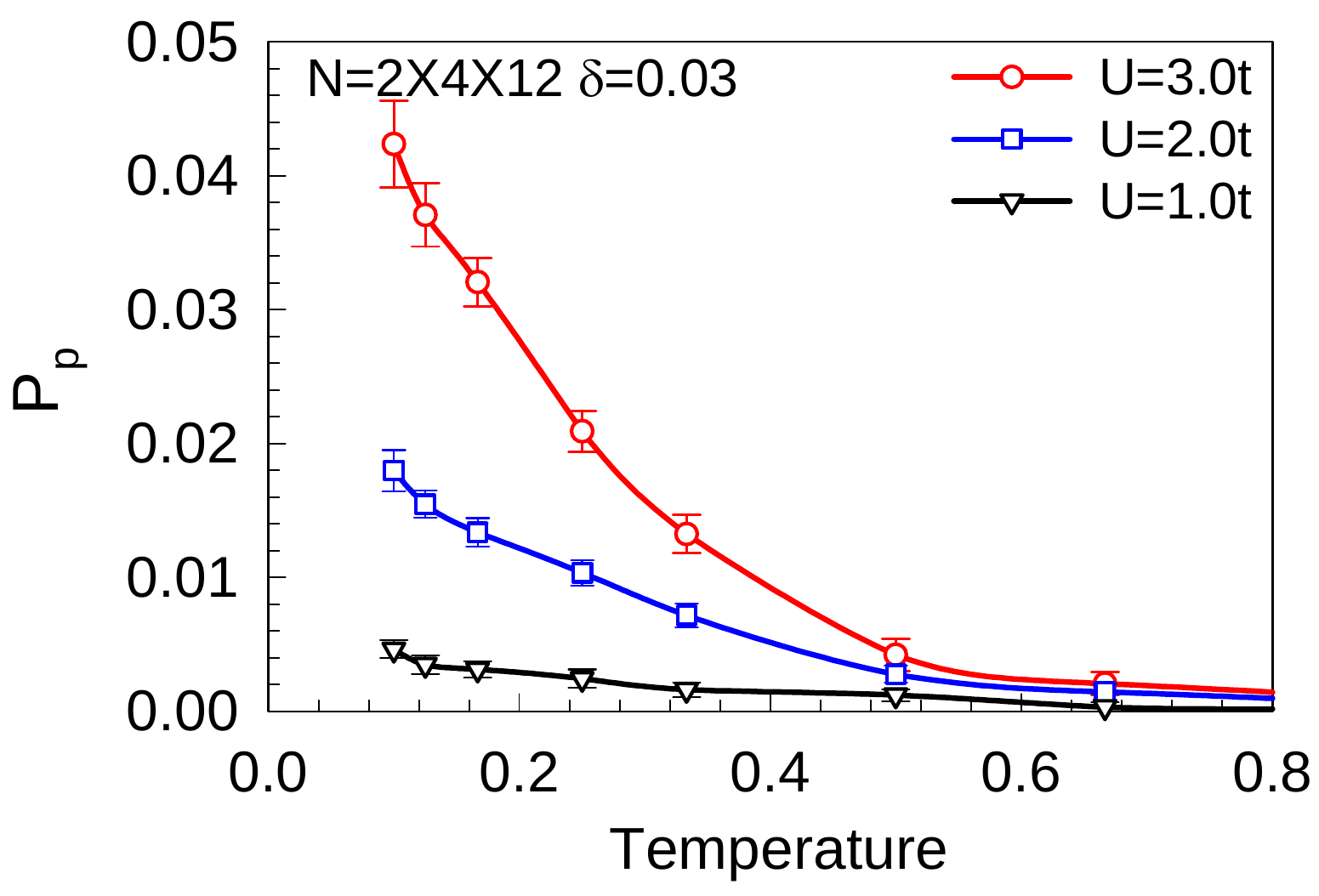}
\caption{(Color online) The effective $p$-type pairing interaction as a function of temperature on a lattice with $2\times 4 \times 12$ sites 
for different $U$ at $\delta=0.03$.
}\label{Fig:Pe}
\end{center}
\end{figure}

In Fig. \ref{Fig:Pe}, we
plot the effective pairing interaction $P_\alpha$ as a function of temperature for different $U$ and
electron fillings on a lattice with $2 \times 4 \times 12$ sites. Clearly in Fig. \ref{Fig:Pe}, the intrinsic pairing interaction $P_\alpha$ is positive and increases with the lowering of temperature. Such a temperature dependence of $P_\alpha$ suggests that effective attractions are generated between electrons and that there is instability towards SC in the system at low temperatures. Moreover, Fig. \ref{Fig:Pe}\ shows that the intrinsic pairing interaction for $p$-wave symmetry is enhanced for larger $U$, indicating that the pairing strength increases with the enhancement of the electron correlations. For another extensive-$s$ pairing symmetry, our DQMC results yield negative intrinsic pairing interactions (not shown here).

Numerical approaches, such as DQMC, however, have their own difficulties as follows: they are typically being limited to small sizes and high temperatures, and
experience the infamous fermion sign problem, which cause exponential growth in the variance of the computed results and
hence an exponential growth in computational time as the lattice size is increased and the temperature is lowered\cite{PhysRevD.24.2278}.
In general, to determine the superconducting pairing symmetry by numerical
calculation for models of finite size, we have to look at the 
distance-dependent pair-correlation function at zero temperature. 
To shed light on this critical issue, it is important to discuss the results obtained by using the CPQMC method\cite{PhysRevLett.74.3652,PhysRevB.55.7464} on a larger lattice. In a variety of benchmarking calculations, the CPQMC method has yielded very accurate results on
the ground-state energy and many other ground-state observables for large systems\cite{PhysRevB.55.7464}.

\begin{figure}[tbp]
\begin{center}
\includegraphics[scale=0.4]{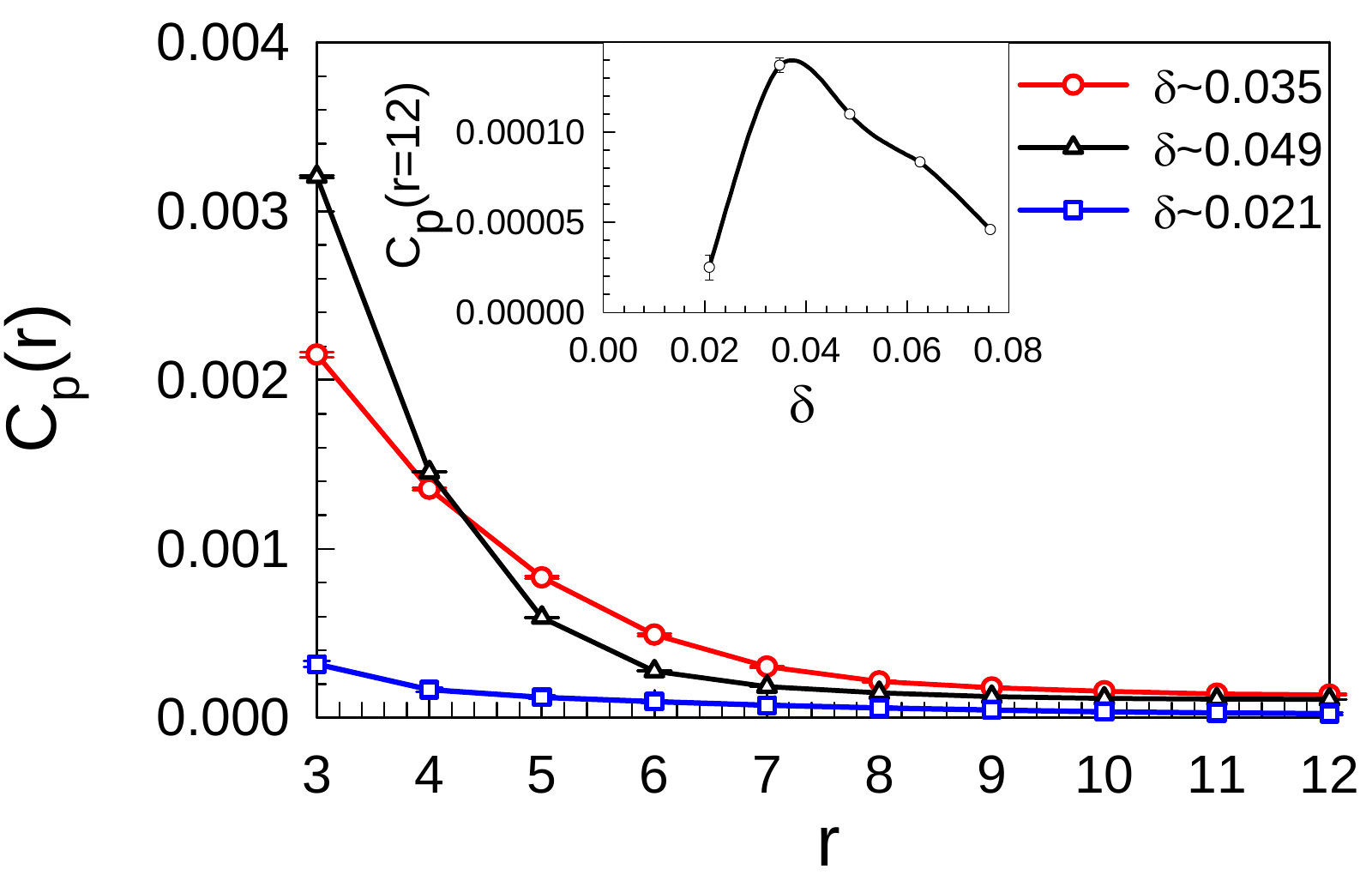}
\caption{(Color online) The $p$-wave superconducting pairing correlation as a function of the distance $r$ on a lattice with $2\times 6 \times 24$ sites.
Inset: the doping-dependent pairing correlation at $r=12$. }
\label{Fig:Pdr}
\end{center}
\end{figure}

In Fig.~\ref{Fig:Pdr}, we compare the pairing correlations on lattices with $2\times 6 \times 24$ sites for different electron fillings at $U=2.0t$. Here, the simulations are
performed for the closed-shell cases. The distance-dependent pairing correlations for $\delta=3/144\simeq 0.021$ (dark triangle), $\delta=5/144\simeq 0.035$ (red circles ), and $\delta=7/144\simeq 0.049$ (blue square) are shown. 
One can readily see that $C_{p}(r)$ decreases as the distance increases, and  
the decay of the distance-dependent pairing correlations is different for different dopings.
In the inset of Fig.\ref{Fig:Pdr}, the pairing correlation $C_{p}(r=12)$ at the largest distance is shown as a function of the doping.
In the filling range that we investigated, $C_{p}(r=12)$ is not a monotonic function of the
doping and there exists an ``optimal" doping (approximately 0.035
electron/site) at which the magnitudes of $C_{p}(r=12)$ are maximized.
This result is consistent with that of RPA, where the doping-dependent pairing correlation bears some similarity to the famous superconducting "dome" in the phase diagram
   of high-temperature superconductors\cite{RevModPhys.78.17}, while the spin correlation at the edge is weakened as the system is doped away from half filling.

\section*{Discussion}

We performed a combined RPA and quantum Monte Carlo study of the magnetic and pairing correlations at the edges in low-doped
zigzag graphene nanoribbons. Our studies show that the triplet edge $p$-wave SC occurs as the ground state of
our model system. The optimal doping is approximately 0.03, which can be easily understood as the DOS peaks at this doping level,
and this doping level is currently achievable experimentally capability for graphene-based material.
Our accurate numerical results establish the properties of the $p$-wave superconducting correlation in zigzag graphene nanoribbons, and will be important for any experimental scheme aimed at detecting the $p$-type superconducting state,
as such a scheme will likely be based on the distinctive properties of the edge.

\section*{Methods}

 The electronic and magnetic properties of the studied system can be well described by the following Hubbard model\cite{RevModPhys.81.109}
\begin{eqnarray}
H&=&-t\sum_{\left\langle i,j\right\rangle}c^{\dagger}_{i\sigma}c_{j\sigma}-t'\sum_{\left\langle\langle i,j\right\rangle\rangle}c^{\dagger}_{i\sigma}c_{j\sigma}+U\sum_{i}n_{i\uparrow}n_{i\downarrow} +\mu\sum_{i\sigma}n_{i\sigma }\label{H}
\end{eqnarray}
where $c^\dag_{i\sigma}$ is the electron-creation operator at site $i$ with spin polarization $\sigma=\A,\V$,
$U$ denotes the on-site repulsive interaction, and $\mu$ is the chemical potential. Here,
the $t$ and $t'$ terms describe the nearest-neighbor (NN) and next nearest-neighbor (NNN) hoppings, respectively.
In the following study, we adopted $t^{\prime}=-0.1t$, which is consistent with experiments\cite{PhysRevB.88.165427}.
In our calculation, we employ periodic boundary conditions in the $x$ direction and open boundary conditions at the zigzag edge.

Specifically, we compute the spin correlation
$S_{i,j}=\langle S_{i}\cdot S_{j}\rangle$ in the $z$ direction, and define the uniform spin susceptibility 
at zero frequency,
\begin{eqnarray}
\chi= \frac{1}{N_s}\int_{0}^{\beta}d\tau \sum_{d,d'=a,b} \sum_{i,j}
 \langle\textrm{m}_{i_{d}}(\tau) \cdot
\textrm{m}_{j_{d'}}(0)\rangle
\end{eqnarray}
To investigate the SC property, we compute the effective pairing
interaction and study the distance dependent pairing correlation.
The effective pairing interaction is extracted from the pairing susceptibility,
\begin{equation}
p_{\alpha}=\frac{1}{N_s}\sum_{i,j}\int_{0}^{\beta }d\tau \langle \Delta
_{\alpha }^{\dagger }(i,\tau)\Delta _{\alpha }^{\phantom{\dagger}%
}(j,0)\rangle,\label{sus}
\end{equation}
with
\begin{eqnarray}
\Delta_{\alpha }^{\dagger }(i)\ =\sum_{l}f_{\alpha}^{\dagger}
(\delta_{l})(c_{{i}\uparrow }c_{{i+\delta_{l}}\downarrow }-
c_{{i}\downarrow}c_{{i+\delta_{l}}\uparrow })^{\dagger}.
\end{eqnarray}
Here, $\alpha$ stands for the pairing symmetry, $f_{\alpha}(\bf{\delta}_{l})$ is the form factor of the pairing
function, and the vectors $\bf{\delta_{l}}$ $(l=1,2)$ denote the
NNN sites along the edge. To extract the effective pairing interaction, the bubble contribution $\widetilde{p}
_{\alpha }(i,j)$ indicating  $\langle
c_{{i}\downarrow }^{\dag }c_{{j}\downarrow }c_{i+\delta_{l}\uparrow}^{\dag}
c_{j+\delta_{l'}\uparrow}\rangle $ in Eq. (\ref{sus}) is being replaced by $\langle c_{{i}\downarrow }^{\dag
}c_{{j}\downarrow }\rangle \langle c_{i+\delta_{l}\uparrow }^{\dag }
c_{j+\delta_{l'}\uparrow }\rangle $. Thus we have obtained an effective pairing interaction $P_\alpha=p_{\alpha}-\widetilde{p}_\alpha$.
The corresponding pairing correlation is defined as
\begin{eqnarray}
C_{\alpha }({\bf{r=R_{i}-R_{j}}})=\langle \Delta _{\alpha }^{\dagger }
(i)\Delta _{\alpha }^{\phantom{\dagger}}(j)\rangle.
\end{eqnarray}
Considering the special structure of
graphene zigzag nanoribbons shown in Fig.\ref{Fig:Structure}, the interesting pairing correlation in such a system is the
pairing between sites on the same sublattice, and two form factors shall be studied
\begin{eqnarray}
\text{$ES$-wave} &\text{:}&\ f_{ES}(\delta_{l})=1,~l=1,2 \\
\text{$p$-wave} &\text{:}&\ f_{p}(\delta_{l})=e^{i(l-1)\pi  },~l=1,2
\end{eqnarray}




\section*{Acknowledgements}
Ma T. thanks CAEP for partial financial support. This work is supported in part by NSCFs (Grant Nos. 11374034, 11274041 and 11334012). 
Yang F. is supported by the NCET program under the grant No. NCET-12-0038.
Lin H.-Q. acknowledges support from NSAF U1530401 and computational resource from the Beijing Computational Science Research Center.
We also acknowledge the support from the HSCC of Beijing Normal University, and the Special Program for Applied Research on Super Computation of the NSFC-Guangdong Joint Fund (the second phase).
\section*{Author contributions statement}

Ma T. provided the QMC data, Yang F. provided the RPA results. Ma T., Yang F, Huang Z., and Lin H.-Q. provided the theoretical understanding and wrote the paper together.

\section*{Additional information}
\subsection*{Competing financial interests}The authors declare no competing financial interests.

\end{document}